\documentstyle[aps,epsfig,multicol]{revtex}

\newcommand{\be}{\begin{equation}}
\newcommand{\ee}{\end{equation}}
\newcommand{\bea}{\begin{eqnarray}}
\newcommand{\eea}{\end{eqnarray}}
\newcommand{\lb}{\left[}
\newcommand{\rb}{\right]}
\newcommand{\lp}{\left(}
\newcommand{\rp}{\right)}

\def\breakon{\end{multicols}\widetext\vspace{-.2cm}
\noindent\rule{.48\linewidth}{.3mm}\rule{.3mm}{.3cm}\vspace{.0cm}}

\def\breakoff{\vspace{-.2cm}
\noindent
\rule{.52\linewidth}{.0mm}\rule[-.27cm]{.3mm}{.3cm}\rule{.48\linewidth}{.3mm}
\vspace{-.3cm}
\begin{multicols}{2}
\narrowtext}

\begin{document}
\draft

\title{Electron shot noise beyond the second moment}
\author{L. S. Levitov$^a$ and M. Reznikov$^b$}
\address{$^a$ Department of Physics, Center for Materials Sciences \& Engineering,\\
Massachusetts Institute of Technology, 77 Massachusetts Ave, Cambridge, MA 02139\\
$^b$ Department of Physics, Solid State Institute, Technion, 32000 Haifa, Israel}

\date{\today}

\maketitle

\begin{abstract}
    The form of electron counting statistics of the tunneling current noise
in a generic many-body interacting electron system is obtained.
The third correlator of current fluctuations
(the skewness of the charge counting distribution) has a universal relation
with the current $I$ and the quasiparticle charge $e^\ast$.
This relation $C_3=(e^\ast)^2I$ holds in a wide bias voltage range, both at
large and small $e V/k_{\rm B}T$, thereby representing an advantage
compared to the Schottky formula.
We consider the possibility of using the counting statistics
for detecting quasiparticle charge at high temperature.
\vskip2mm
\end{abstract}

\bigskip

\begin{multicols}{2}
\narrowtext


Recent developments in the problem of quantum electron transport
were marked by interest in the phenomenon of electric noise.
The many-body theory of electron shot noise, developed
by Lesovik \cite{lesovik89} (and independently
by Khlus \cite{khlus87}) for a point contact, was extended
to multiterminal systems by B\"uttiker \cite{buttiker90} and
to mesoscopic systems by Beenakker and B\"uttiker \cite{beenakker92}.
Kane and Fisher proposed using shot noise for detecting
fractional quasiparticles in a Quantum Hall Luttinger liquid
\cite{kanefisher93}.

Experimental studies of the shot noise, after first measurements
in a point contact
by Reznikov et al. \cite{reznikov95} and Kumar et al. \cite{kumar96},
focused on the quantum Hall regime. The fractional charges $e/3$ and $e/5$
were observed \cite{depicciotto97,glattli97,reznikov99} at
incompressible Landau level filling (see also recent
work on noise at intermediate filling \cite{heiblum}).
The shot noise in a mesoscopic conductor was observed by
Steinbach et al. \cite{steinbach96} and Schoelkopf et al. \cite{schoelkopf97},
who also studied noise in an ac driven
phase-coherent mesoscopic conductor \cite{schoelkopf98}.

In this article we discuss a generalization of the shot noise, namely
the counting statistics of fluctuating electric current.
It can be defined
through the probability distribution $P(q)$ of charge transmitted
in a fixed time interval \cite{levitov93,levitov96}.
We consider ways of obtaining the distribution $P(q)$
using a fast charge integrator scheme.
From the distribution $P(q)$ all moments of
charge fluctuations can be calculated and,
conversely, the knowledge of all moments is in principle
sufficient for reconstruction of the full distribution.
However, due to the central limit theorem,
high moments are difficult to access experimentally.
Therefore we shall focus primarily on the third moment.

The counting statistics have been analyzed theoretically
for a Fermi gas, in the single- and multi-channel geometry
\cite{levitov93,ivanov93}, in the mesoscopic regime \cite{hwlee95,nazarov},
and in the ac driven phase-coherent regime \cite{ivanov93,ivanov97}.
Charge doubling
due to Andreev scattering in NS junctions was considered
by Muzykantskii and Khmelnitskii \cite{muzykantskii},
and in mesoscopic NS systems by Belzig and Nazarov \cite{nazarovNS}.
However, since the most interesting applications of the shot noise lie
in the domain of interacting electron systems, an appropriate extension
of the theory is necessary.

The problem of back influence of a charge detector on current fluctuations
was considered by Lesovik and Loosen \cite{lesovik97},
and recently by Nazarov and Kindermann \cite{nazarov-general}.
Beenakker proposed
an alternative way of obtaining charge statistics using
photon counting \cite{beenakker-photons}. 
Application to pumping in quantum dots was 
also discussed \cite{AndreevKamenev}.

Our central finding is a relation between counting statistics
and the Kubo theorem, valid {\it in the tunneling regime} for a generic
interacting many-body system. From that we obtain
a formula for the moments of the counting statistics
that holds in the entire bias voltage range,
at arbitrary $eV/k_{\rm B}T$.
We demonstrate that in the tunneling regime the current fluctuations
are described by
an uncorrelated mixture of two Poisson processes.
This is revealed by a generating function
$\chi(\lambda)=\sum_{q} P(q) e^{i\lambda q/e^\ast}$, with
$e^\ast$ the quasiparticle charge. We find
  \be\label{eq:2poisson}
\chi(\lambda)=
\exp\lb (e^{i\lambda}\!-\!1)N_{1\to 2}(\tau)+(e^{-i\lambda}\!-\!1)N_{2\to 1}(\tau)\rb
  \ee
where $N_{a\to b}(\tau)=m_{ab}\tau$ is the mean charge number
transmitted from the contact $a$ to the contact $b$ in a time $\tau$.

\begin{figure}
\centerline{\psfig{file=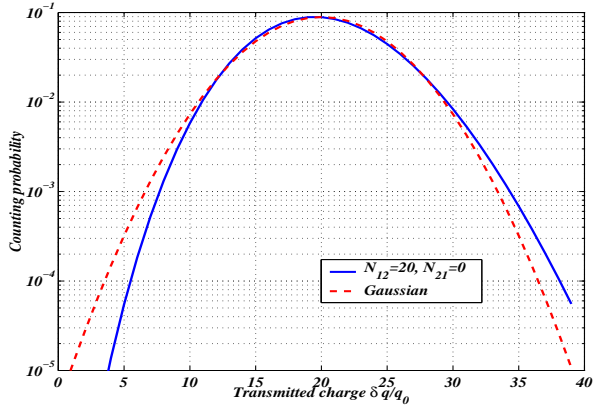,width=3.1in,height=2.15in}}
\vspace{0.5cm}
        \caption[]{
The third moment (\ref{eq:q3ave}) determines
the shape of 
the distribution $P(q)$, namely its {\it skewness}. 
This is illustrated by
a distribution of the form (\ref{eq:2poisson})
and a Gaussian with the same mean and variance. 
For $C_3>0$
the peak is somewhat more stretched to the right than to the left.
        }
\label{fig:P(q)}
\end{figure}

The result (\ref{eq:2poisson}) yields a number of relations between
different statistics of the probability distribution $P(q)$.
The cummulants
$\langle\!\langle\delta q^k\rangle\!\rangle $ (irreducible correlators)
of the distribution
$P(q)$ are expressed
in terms of $\chi(\lambda)$ as
  \be
\ln\chi(\lambda)=
\sum_{k=1}^{\infty}\frac{(i\lambda)^k}{k!}
\frac{\langle\!\langle\delta q^k\rangle\!\rangle }{(e^\ast)^k}
  \ee
Using Eq. (\ref{eq:2poisson}) one obtains
  \be\label{eq:qmoments}
\langle\!\langle\delta q^k\rangle\!\rangle =\, (e^\ast)^k \,\cases{(m_{12}-m_{21})\tau,\quad & $k$ odd\cr
(m_{12}+m_{21})\tau,\quad & $k$ even}
  \ee
Setting $k=1,2$ we express $m_{12}\pm m_{21}$ through the time-averaged current
and the low frequency noise power:
  \be\label{eq:IP}
m_{12}-m_{21}=I/e^\ast, \quad
m_{12}+m_{21}= {\cal P}/2(e^\ast)^2.
  \ee
Of special interest for us will be the cummulant 
$\langle\!\langle\delta q^3\rangle\!\rangle$ which is equal 
to the third correlator \cite{corr4}
  \be\label{eq:q3ave}
\langle\!\langle\delta q^3\rangle\!\rangle
 \equiv \overline {\delta q^3}=\overline{\lp q-\overline{q}\rp^3}
  \ee
(see Fig. \ref{fig:P(q)}).
For this correlator
Eq. (\ref{eq:qmoments}) gives
$\langle\!\langle\delta q^3\rangle\!\rangle =C_3\tau$ with the
coefficient $C_3$
(``spectral power'') related to the
current $I$ as
  \be\label{eq:q3-I}
C_3\equiv \langle\!\langle\delta q^3\rangle\!\rangle /\tau =(e^\ast)^2 I
  \ee
We note that the relation (\ref{eq:q3-I}) holds for the distribution
(\ref{eq:2poisson})
at any ratio of the mean number of transmitted charges $m_{12}-m_{21}$
to the variance $m_{12}+m_{21}$.

The meaning of Eq. (\ref{eq:q3-I}) is similar to that of the Schottky formula
for the second correlator
${\cal P}=2\langle\!\langle\delta q^2\rangle\!\rangle =2e^\ast I$
which is usually used to
determine the effective charge $e^\ast$ from the tunneling current noise.
The Schottky formula is valid when charge flow is unidirectional,
which means $m_{12}\gg m_{21}$ (see Eq. (\ref{eq:qmoments})).
The latter can be true
only at sufficiently low temperatures $k_{\rm B}T\ll eV$. This requirement
of a cold sample at a relatively high bias voltage
is the origin of a well known difficulty in the noise measurement.
In contrast, the relation (\ref{eq:q3-I}) is not constrained by
any requirement on sample temperature.

On a general basis we expect the relation (\ref{eq:q3-I}
to hold approximately even outside the tunneling regime. 
Indeed, for the Nyquist noise at equilibrium all odd moments vanish.
Combined with the temperature independence of $P(q)$
out of equilibrium, this implies a relatively weaker dependence
on $eV/k_{\rm B}T$ than in the noise power at
the Nyquist-Schottky crossover.
This is manifest,
for instance, in the temperature independent first moment of $P(q)$
for free fermions (the Landauer formula).

This property of the third moment, if confirmed experimentally,
may prove to be quite useful for determining the quasiparticle
charge. In particular, this applies to the situations when the
$I-V$ characteristic is strongly nonlinear, when it is usually
difficult to unambiguously interpret the variation of the second
moment with current as a shot noise effect or as a result
of thermal noise modified by non-linear conductance.
We stress that this is a completely general problem
pertinent to any interacting system. Namely, in systems such as
Luttinger liquids, the $I-V$ nonlinearities arise at $eV\ge k_{\rm B}T$.
However, it is exactly this voltage that has to be applied for
measuring the shot noise in the Schottky regime.


Now we turn to the derivation of the main result (\ref{eq:2poisson}).
The starting point of our analysis will be the tunneling Hamiltonian
$\hat{\cal H}=\hat{\cal H}_1+\hat{\cal H}_2+\hat V$,
where $\hat{\cal H}_{1,2}$ describe the
leads and $\hat V=\hat J_{12}+\hat J_{21}$ is the tunneling operator.
The specific form of the operators $\hat J_{12}$, $\hat J_{21}$
that describe tunneling of a quasiparticle
between the leads will be unimportant for the most of our discussion.

The counting statistics generating function $\chi(\lambda)$ can be written
\cite{levitov96} as a Keldysh partition function
  \be\label{eq:Zkeldysh}
\chi(\lambda) = \Big\langle {\rm T}_{\rm K}
\exp\lp -i\int_{C_{0,\tau}}\hat{\cal H}_{\lambda}(t)dt\rp\Big\rangle
,
  \ee
where a counting field $\lambda(t)$ is added to the phase of
the tunneling operators $\hat J_{12}$, $\hat J_{21}$ as
  \be\label{Vlambda}
\hat V_{\lambda}= e^{\frac{i}2\lambda(t)}\hat J_{12}(t)
+ e^{-\frac{i}2\lambda(t)}\hat J_{21}(t)
  \ee
Here $\lambda(t)=\pm\lambda$ is antisymmetric on the forward and backward
parts of the Keldysh contour $C_{0,\tau}\equiv \lb 0\to \tau\to 0\rb$.
Eqs. (\ref{eq:Zkeldysh}), (\ref{Vlambda}) originate from the analysis
of a coupling Hamiltonian for an ideal ``passive charge detector'' without
internal dynamics \cite{levitov96,nazarov-general}.

In what follows we compute $\chi(\lambda)$ and establish a relation with
the Kubo theorem for tunneling current \cite{mahan}.
For that, we perform the usual
gauge transformation turning the bias voltage into the tunneling
operator phase factor as
$\hat J_{12}\rightarrow \hat J_{12}e^{-ieVt}$,
$\hat J_{21}\rightarrow \hat J_{21}e^{ieVt}$.
Passing to the Keldysh interaction representation, we write
  \be\label{eq:chi-V}
\chi(\lambda) = \Big\langle {\rm T}_{\rm K}
\exp\lp -i\int_{C_{0,\tau}}\hat V_{\lambda(t)}(t)dt\rp\Big\rangle
  \ee
Diagrammatically, the partition function (\ref{eq:chi-V}) is a sum
of linked cluster diagrams with appropriate combinatorial factors.
To the lowest order in the tunneling operators $\hat J_{12}$, $\hat J_{21}$
we only need to consider linked clusters of the second order.
This gives $\chi(\lambda) =e^{W(\lambda)}$, where
  \be
W(\lambda)=-\frac12
\int\!\!\int_{C_{0,\tau}}
\!\Big\langle {\rm T}_{\rm K} \hat V_{\lambda(t)}(t)
\hat V_{\lambda(t')}(t')\Big\rangle\,
dtdt'
  \ee
There are several different contributions to this integral, from $t$ and $t'$
on the forward or backward
parts of the contour $C_{0,\tau}$. Evaluating them separately, we obtain
  \bea\label{Wlambda2}
&& W(\lambda)=
\int_0^\tau\!\!\int_0^\tau\!\Big\langle \hat V_{-\lambda}(t)
\hat V_{\lambda}(t')\Big\rangle\,
dt'dt
\\ \nonumber
&& -\!\int_0^\tau\!\!\int_0^{t}\!\!\Big\langle\! \hat V_{\lambda}(t)
\hat V_{\lambda}(t')\!\Big\rangle\,
dt'\!dt
-\!\int_0^\tau\!\!\int_{t}^\tau\!\!\Big\langle\! \hat V_{-\lambda}(t)
\hat V_{-\lambda}(t')\!\Big\rangle\,
dt'\!dt
  \eea
We substitute the form (\ref{Vlambda}) into Eq.(\ref{Wlambda2})
and average by pairing $\hat J_{12}$ with $\hat J_{21}$. This gives
\bea\label{eq:Wlambda}
&& W(\lambda)=
(e^{i\lambda}\!-\!1)N_{1\to 2}(\tau)+(e^{-i\lambda}\!-\!1)N_{2\to 1}(\tau)
\\\label{eq:N12}
&& {\rm with}\quad N_{a\to b}=
\int_0^\tau\!\!\int_0^\tau \langle \hat J_{ba}(t)\hat J_{ab}(t')\rangle\,dtdt'
\eea
Exponentiating (\ref{eq:Wlambda}) gives the 
result (\ref{eq:2poisson})

It is instructive to relate the quantities (\ref{eq:N12})
with the Kubo theorem.
We consider the tunneling current operator
$\hat{\cal I}(t)=-ie^\ast\lp \hat J_{12}(t)-\hat J_{21}(t)\rp$.
From the Kubo theorem for the tunneling current \cite{mahan},
the mean integrated current
$\int_0^\tau\langle \hat{\cal I}(t)\rangle dt$
scaled by $e^\ast$ is nothing but
  \be\label{Qave}
\int_0^\tau\!\!\int_0^\tau \Big\langle \lb \hat J_{21}(t),\hat J_{12}(t')\rb \Big\rangle\,dtdt'
= N_{1\to 2}-N_{2\to 1}
  \ee
By writing $N_{a\to b}=m_{ab}\tau$, we obtain the first relation (\ref{eq:IP}).
To obtain the second relation (\ref{eq:IP}) we consider the variance
of the charge transmitted in time $\tau$. It is given by a time integral
of an averaged symmetrized product of two current operators \cite{sukhorukov}
  \be\label{Q2ave}
\langle\!\langle\delta q^2\rangle\!\rangle =(e^\ast)^2\int_0^\tau\!\!\int_0^\tau \Big\langle \left\{ \hat{\cal I}(t),\hat{\cal I}(t')\right\}_+ \Big\rangle\,dtdt'
  \ee
The integral in (\ref{Q2ave}) can be rewritten as
  \be
\int_0^\tau\!\!\int_0^\tau\! \Big\langle\! \left\{ \hat J_{12}(t),\hat J_{21}(t')\right\}_+\! \Big\rangle\,dtdt'
= N_{1\to 2}\!+\! N_{2\to 1}
  \ee
which immediately leads to (\ref{eq:IP}).

The result (\ref{eq:2poisson}), and thereby
the formula (\ref{eq:q3-I}) for the 3-rd correlator, are valid only
at low transmission. In that the situation is similar
to the Kubo formula for the tunneling current, which is valid only
in the tunneling Hamiltonian approximation. To illustrate this
we recall the expression for
counting statistics for a single channel noninteracting Fermi system
(point contact) in the presence of a dc voltage $V$ and temperature $T$
\cite{levitov96},
  \bea\label{eq:chi-tr}
&& \chi(\lambda)=\exp\lp -N_T {\cal U}_+{\cal U}_-\rp
,\quad
N_T=\frac{\tau k_{\rm B}T}{2\pi\hbar}
\\
&&{\cal U}_\pm\! ={\cal U}/2 \pm \cosh^{-1}\!\lp t\cosh({\cal U}/2+\! i\lambda)+\!r \cosh \lp{\cal U}/2\rp\rp
  \eea
where $\tau$ is a measurement time, and
${\cal U}=eV/k_{\rm B}T$. This result holds for any values of the
transmission and reflection constants $t$ and $r$ (constrained by $t+r=1$).
The formula (\ref{eq:chi-tr}) was obtained in Ref. \cite{levitov96}
by explicitly evaluating the Keldysh partition function in the
scattering basis representation.

\begin{figure}
\centerline{\psfig{file=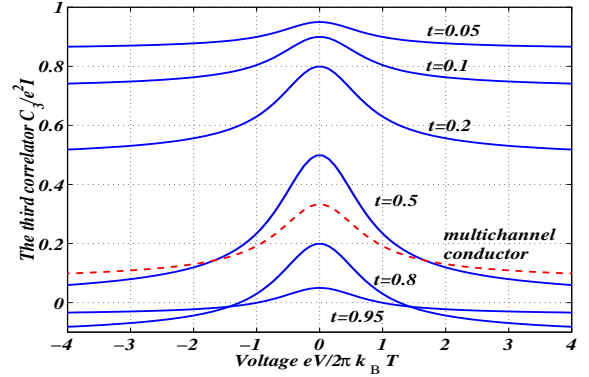,width=3in,height=2in}}
\vspace{0.5cm}
        \caption[]{
The third correlator 
$C_3$ scaled by $e^2I$, with $I$ the time-averaged current 
[see Eqs. (\ref{eq:q3ave}), (\ref{eq:q3-I})],
for the single channel problem 
(\ref{eq:chi-tr}), (\ref{eq:q3-tr}), and for a phase-coherent
mesoscopic multichannel conductor. Note that the 
relation $C_3=e^2I$ holds approximately at not too large 
transmission $t$. 
        }
\label{fig:C3-tr}
\end{figure}

The 3-rd correlator $\langle\!\langle\delta q^3\rangle\!\rangle $ can be obtained from
(\ref{eq:chi-tr}) by expanding $\ln\chi(\lambda)$ in Taylor series
up to $O(\lambda^3)$:
  \be\label{eq:q3-tr}
\langle\!\langle\delta q^3 \rangle\!\rangle 
= e^3 t(1-t)N_T
\lp 6t\frac{\sinh{\cal U}-{\cal U}}{\cosh{\cal U}-1} +(1-2t){\cal U}\rp
  \ee
This expression is a function of the bias-to-temperature ratio ${\cal U}$,
and so in this case the relation (\ref{eq:q3-tr}) for the 3-rd correlator
does not hold (see Fig. \ref{fig:C3-tr}). Asymptotically
  \be\label{eq:q3-tr-cases}
\langle\!\langle\delta q^3\rangle\!\rangle =\cases{e^2(1-t)I\tau,\quad &$eV\ll k_{\rm B}T$\cr
e^2(1-2t)(1-t)I\tau,\quad &$eV\gg k_{\rm B}T$}
  \ee
where $I=\frac{e^2}{2\pi\hbar}tV$. One can also average over
the universal Dorokhov's distribution of transmission
in a multichannel mesoscopic metal \cite{hwlee95} (see Fig. \ref{fig:C3-tr}).

Eqs. (\ref{eq:q3-tr}), (\ref{eq:q3-tr-cases}) indicate nonuniversality
of the relation (\ref{eq:q3-I})
outside the tunneling regime. They also lead to an interesting
qualitative prediction: At $t>0.5$ the ratio
$\langle\!\langle\delta q^3\rangle\!\rangle/I$ can become negative.
Such a signature could be observed even if it proves difficult
to measure $\langle\!\langle\delta q^3\rangle\!\rangle$
quantitatively with sufficient precision.
This is important in view of the difficulties
in measuring the counting statistics (see below).

In the single channel problem (\ref{eq:chi-tr})
the tunneling regime is realized
at low transmission $t$. To connect
with the results
(\ref{eq:2poisson}), (\ref{eq:q3-I})
we analyze the expression (\ref{eq:chi-tr})
at $t\ll 1$.
To the lowest order in small $t$ we have
  \be
{\cal U}_+={\cal U}, \quad
{\cal U}_-=t\frac{e^{{\cal U}}\lp e^{i\lambda}-1\rp
+\lp e^{-i\lambda}-1\rp}{e^{\cal U}-1}
  \ee
Substituting this in Eq.(\ref{eq:chi-tr}) we recover
(\ref{eq:2poisson}) with
  \be
N_{2\to 1}(\tau)= \frac{eV\tau}{2\pi\hbar}\frac{t}{e^{\cal U}-1}
,\quad
N_{1\to 2}(\tau)=e^{\cal U} N_{2\to 1}(\tau)
  \ee
the rates of two Poisson processes.

The measurement of the distribution $P(q)$ is a nontrivial task.
Current fluctuations must be amplified with a very low noise
preamplifier ({\em e.g.} the one used in Refs.
\cite{depicciotto97,reznikov99}). Amplified signal can then be
digitized and analyzed with computer. This setup in principle allows
to reconstruct the full statistics of transmitted charge. In practice, however,
the correlators of high order become increasingly difficult to extract.

The main source of error in the measurement of 
the $k$-th cummulant $C_k$ of the distribution $P(q)$ is statistical. 
The nongaussian character of the amplifier noise
does not present a problem, since the mean time-averaged value of the 
$k$-th cummulant $C_k^{(a)}$ for the amplifier can be subtracted
if known with sufficient accuracy. 
The measured value
$C_k$ should be compared to (i) the variance 
${\rm var\ }C_k$ of the $k$-th cummulant statistics and
(ii) the variance ${\rm var\ } C_k^{(a)}$
of the $k$-th cummulant of the amplifier noise. The variance 
is in both cases expressed through the 
correlators of order $2k$. The correlators of {\it even} order
for a generic distribution 
can be estimated, by virtue of 
the central limit theorem, using Gaussian statistics:
  \be \label{eq:deltak}
 {\rm var\ }C_k = \lp\langle \delta
q^{2k}\rangle\rp^{1/2}\simeq \lp (2k-1)!!\, \langle
 \delta q^2 \rangle^k \rp^{1/2}
 \ee
   Fluctuations
introduced by amplifier can also be estimated using Gaussian statistics. 
For the amplifier noise
of spectral density $A$ (measured in ${\rm A^2/Hz}$), charge fluctuations
are $\delta Q^2=\frac{1}{2} A\tau$, where $\tau$ is sampling time. An estimate 
of the variance ${\rm var\ } C_k^{(a)}$, similar
to Eq.(\ref{eq:deltak}), gives
  \be \label{eq:Deltak}
{\rm var\ } C_k^{(a)}
=\lp (2k-1)!! \lp {\textstyle \frac{1}{2}} A\tau\rp^k \rp^{1/2}
  \ee
For odd $k$ the commulant 
$C_k=\langle\!\langle \delta q^k \rangle\!\rangle$ mean value is
proportional to the current $I$, as discussed above. 
The fluctuations due to the amplifier
are independent of $I$. In the Nyquist regime (at small $I$) 
the variance ${\rm var\ } C_k$
is independent on $I$ and is determined by thermal noise. 
In the Schottky noise regime (at large $I$) the fluctuations 
are determined by $I$, and ${\rm var\ } C_k\propto I^{k/2}$.
Therefore at $k>2$ one can increase the signal-to-noise
ratio by increasing the current until 
${\rm var\ } C_k \approx {\rm var\ } C_k^{(a)}$

In the shot noise regime, when
$\langle \delta q^2 \rangle
= eI\tau \approx A\tau$, we can
estimate the signal-to-noise ratio as
  \be \label{eq:sn1} 
S/N\simeq \frac{\langle\!\langle
q^k\rangle\!\rangle}{2 {\rm var\ } C_k}
\simeq\frac{1}{2((2k-1)!!)^{1/2}}\left(\frac{2e^2}{A\tau}\right)^{k/2-1}
  \ee
It is clear from Eq.(\ref{eq:sn1}) that it is beneficial to decrease
the sampling time $\tau$ to gain sensitivity. Estimates for typical
values of $A=10^{-28}~{\rm A^2/Hz}$ and $\tau=10^{-7}~{\rm
s^{-1}}$ give for the $k=3$ commulant $S/N\approx 10^{-2}$. 
This value is acceptable, since repeating the measurement many times 
over a long time $\cal T$
and averaging will further reduce
statistical fluctuations by a factor of $\sqrt{\cal T/\tau}$.
For the cummulants of higher order $k>3$ the situation is 
more problematic.

In summary, the counting statistics (\ref{eq:2poisson}) of
tunneling current is found to be universal and independent of the
character of interactions. For the third correlator we obtain a
generalized Schottky formula (\ref{eq:q3-I}). This formula is
valid at both large and small $eV/k_{\rm B}T$ and can be used to
measure quasiparticle charge at temperatures $k_{\rm B}T\ge eV$.
A method for measuring the third correlator is proposed.

This research is supported by the Binational Science Foundation
and by the MRSEC Program of the National Science Foundation under
Grant No. DMR 98-08941.

\end{multicols}

\begin{thebibliography}{99}
\bibitem{corr4}  
The relation between cummulants and correlators is generally 
more complicated than Eq.(\ref{eq:q3ave}) for the third cummulant.
For example, 
$\langle\!\langle\delta q^4\rangle\!\rangle
= \overline {\delta q^4}-3\lp \overline {\delta q^2} \rp^2$.
\end{thebibliography}
\end{document}